\documentclass[journal]{IEEEtran}
\usepackage{multicol}

\usepackage{cite}

\ifCLASSINFOpdf

\else

\fi

\usepackage[cmex10]{amsmath}
\usepackage[cmex10]{amsmath}
\usepackage{amssymb}

\usepackage{array}
\usepackage{mdwmath}
\usepackage{mdwtab}

\newcommand{\ie}{\emph{i.e. }}

\newcommand{\m}[1]{\mathcal{#1}}

\usepackage[dvips]{graphicx}
\graphicspath{{eps/}}
\DeclareGraphicsExtensions{.eps}

\IEEEoverridecommandlockouts

\begin{document}
%
\title{Design of Low Complexity Non-binary LDPC Codes with an Approximated Performance-Complexity Tradeoff}

\author{{Yang Yu, and Wen~Chen,~\IEEEmembership{Senior Member,~IEEE}}
\thanks{Manuscript received December 6th, 2011; revised February 1st, 2012, accepted
February 8th, 2012. The associate editor coordinating the review of
this paper and approving it for publication was M. Lentmaier.}
\thanks{The authors are with
Department of Electronic Engineering, Shanghai Jiao Tong University,
Shanghai, and SKL for ISN, Xidian University, China. e-mail:
\{yuyang83;wenchen\}@sjtu.edu.cn.}
\thanks{This work is supported by NSFC \#60972031, by national 973 project
\#2012CB316106 and \#2009CB824900, by NSFC \#61161130529, by
national key laboratory project \#ISN11-01.}}

\markboth{IEEE Communications Letters. VOL. X. NO. X. XXXXXX 2012}%
{\MakeLowercase{Design of Low Complexity Nonbinary LDPC Codes with an Approximated Performance-Complexity Tradeoff}}

\maketitle

\begin{abstract}
By presenting an approximated performance-complexity tradeoff (PCT) algorithm,
a low-complexity non-binary low density parity check (LDPC) code
over $q$-ary-input symmetric-output channel
is designed in this manuscript
which converges faster than the threshold-optimized non-binary LDPC codes
in the low error rate regime.
We examine our algorithm by both hard and soft decision decoders.
Moreover, simulation shows that the approximated PCT algorithm
has accelerated the convergence process by $30\%$ regarding the number of the decoding iterations.

\end{abstract}


%
\IEEEpeerreviewmaketitle

\section{Introduction}
Investigation over Galois field $GF(q)$, $q=2^p$,
shows that $q$-ary LDPC codes have potentially better performance than binary LDPC codes for not very long block length
 at the cost of higher decoding complexity, and irregular LDPC codes can outperform the regular LDPC codes\cite{Davey98montecarlo}. The design of high-performance nonbinary LDPC codes has been studied in the literature\cite{Byers05ExitNonbinary,Rathi05DeNonbinary,ge09NLDPC}.
A major concern of $q$-ary LDPC is the decoding complexity.


PCT analysis in \cite{Wei05PctB,smith10LdpcPCT} utilizes the nature of binary iterative decoder, in which messages passing through each iteration, can be profiled by a single parameter.
The code design problem is then reduced to the shaping of the decoding trajectory of extrinsic information transfer (EXIT) chart for an optimal PCT\cite{smith10LdpcPCT}, where they show that the (decoding) complexity optimized binary LDPC codes outperforms the threshold optimized binary LDPC codes. However, messages, passing through the nonbinary LDPC decoder, are vectors\cite{ge09NLDPC}. The main challenge in cooperating PCT in nonbinary LDPC codes design is how to characterize the decoding complexity as a uni-parametric transfer function. To solve this problem, we present an irregular EXIT chart by using an upper bound of the message error probability, based on which, an approximated performance-complexity tradeoff (PCT) algorithm is put forward to design irregular nonbinary LDPC codes with optimized decoding complexity.


However, advantages of the proposed approximated PCT algorithm are obvious:
firstly, \cite{Byers05ExitNonbinary,Rathi05DeNonbinary} give methods to predict the performance threshold for nonbinary LDPC codes, but the complexity can not be optimized based on these procedures. The presented EXIT chart
can be also used to reduce the decoding complexity.
Secondly, the complexity optimization algorithm in \cite{Wei05PctB,smith10LdpcPCT} is applicable for binary LDPC codes with rates greater than $1/4$. But the optimization algorithm in this manuscript is a universal method in the sense that, when $q=2$, the algorithm coincides with binary case.

\section{Preliminaries}
\subsection{LDPC Codes}
\begin{figure}[!hbtp]
\begin{center}
\includegraphics[width=1.6in]{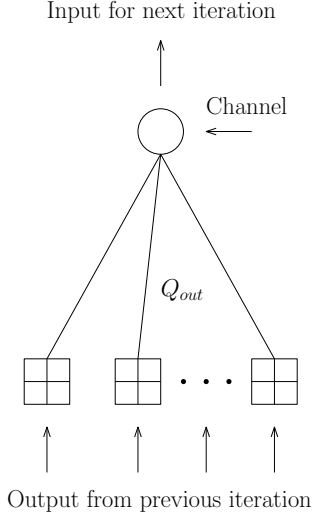}
\end{center}
\vspace*{-.3cm}
\caption{Depth-one decoding tree.}
\label{fig:one_tree}\vspace*{-.3cm}
\end{figure}

An LDPC code is called regular if the column and row weight of the parity check
matrix is constant, respectively.
The irregular LDPC codes can be characterized by variable degree distribution
\begin{equation}
  \lambda (x) = \sum \limits_{i \geqslant 2}\lambda_i x^{i-1},
\end{equation}
and check degree distribution
\begin{equation}
  \rho (x) = \sum \limits_{i \geqslant 2}\rho_i x^{i-1},
\end{equation}
from an edge-perspective, where $\lambda_i$ and $\rho_i$ are the fraction of edges belonging to degree-$i$ variable and check node, respectively. Using this characterization, code rate $R$ is given by
$R = 1 - \frac{\int_0^1 \rho (x)dx}{\int_0^1 \lambda (x)dx}$,
and $\lambda(1)=\rho(1)=1$. Due to this characterization, Fig.~\ref{fig:one_tree} gives the depth-one decoding tree for a degree-$i$ variable node. During one iteration, messages (beliefs) are passed from the input to the output of the tree.
The EXIT chart based on message error probability of LDPC codes can be given by
\begin{equation}
\label{p_out}
  p_{out} = \sum \limits_{i \geqslant 2} \lambda_i f_i(p_{in}),
\end{equation}
where $p_{in}$ is the input error probability and $f_i$ is the elementary EXIT chart associated with degree-$i$ depth-one tree\cite{ardakani04EXIT} as in Fig.~\ref{fig:one_tree}. The initial probability is calculated by $p_0 = {\bf{P_e}}(D_{ch})$, where $\bf{P_e}$ denotes the error probability and $D_{ch}$ is the conditional probability distribution function (pdf) of the message from channel. Then the number of decoding iterations is given\cite{Wei05PctB,smith10LdpcPCT} by
\begin{equation}
  N = \int_{p_t}^{p_0}{\left( p\ln \left( \frac{p}{\sum_{i \geqslant 2} \lambda_i f_i(p)} \right)\right)^{-1}dp},
\end{equation}
where $p_t$ is the target error probability.

\subsection{Symmetric Conditions for Nonbinary LDPC Codes}
A log-domain FFT-QSPA (The fast Fourier transform q-ary sum-product algorithm) decoder is used in\cite{ge09NLDPC}.
The log likelihood ratio vectors (LLRV) are fed into the decoder.
The $i$th element of LLRV can be calculated as $l_i = \ln \frac{p_0}{p_i}$,
where $p_i = {\bf Pr}({\bf y}|x=i)$, and ${\bf y}$ is the channel observation of variable node $x$.
In\cite{ge09NLDPC},
a generalized symmetric condition for $q$-ary-input symmetric-output channel ($q$-ary PSK modulated channels for prime $q$ and binary-modulated channels for $q=2^p$) is given by
\begin{equation}
\nonumber
  {\bf Pr}({\bf y}|x=a) = {\bf Pr}({\m I}[a]{\bf y}|x=0), \forall a \in GF(q),
\end{equation}
where ${\m I}[a]$ is a $(q-1) \times (q-1)$ diagonal matrix with the $i$-th diagonal entry $r^{i\otimes a}$, $i = 1,2,...,q-1$, $r$ is the primitive
root of the corresponding field, and $\otimes$ is the mod-$q$ multiplication. Further, it's
proven that, under this symmetric condition, the error performance of an LDPC code
is independent of the transmitted codeword.
So, analysis (EXIT chart) for $q$-ary LDPC codes based on all-zero
codeword will suffice for the decoder.

\section{Complexity-Optimized Nonbinary LDPC Codes}
This section proposes a
irregular nonbinary EXIT chart based on an upper bound of message error probability.
Further, a complexity optimization algorithm
based on the EXIT chart is put forward to design low decoding complexity $q$-ary LDPC codes which are examined by both hard and soft decision decoders.
\subsection{Irregular EXIT Chart for Nonbinary LDPC Codes}
\label{section_EXIT}
Assuming all zero codewords are sent, a well designed EXIT chart can be adopted to construct
$q$-ary LDPC codes with optimized PCT over
$q$-ary-input symmetric channel.
Based on symmetric conditions, EXIT chart is first developed for Turbo codes as pictorial demonstration of iterative decoding process\cite{Brink99EXIT}. Later, a more accurate approximation is applied to binary LDPC to design good performance code ensemble according to their degree distributions\cite{ardakani04EXIT}.
When it is applied to $q$-ary LDPC, \cite{ge09NLDPC} generalizes the symmetric condition, gives a Gaussian approximation to non-binary density evolution, and shows that, by using a channel adapter, static channel can be forced to be symmetric.
A more systematic approach to design $q$-ary LDPC codes is given
in\cite{Bennatan06NonbinaryMem}, where they use
coset codes to symmetrize the memoryless channels,
and design good coset $GF(q)$ LDPC codes too. An EXIT chart based on new mutual information metric is given in\cite{Byers05ExitNonbinary} using a Gaussian mixture distribution which is less computationally intensive. The EXIT chart for $q$-ary LDPC is also studied in\cite{Rathi05DeNonbinary}.

These methods can well predict the performance thresholds of LDPC codes with
infinite block length. But the decoding complexity can not be optimized based on these design procedure. So, instead of giving method for
predicting the precise performance of $q$-ary LDPC codes, we present a
complexity optimization algorithm by using  Gallager's formula
which is an upper bound of message
error probability for FFT-QSPA decoder and can be also used as an extended analysis
for Gallager decoding algorithm b (Gal-b) \cite{Gallager63low-densityparity-check}.

The reasons why we adopt the Gallager's formula to extend the PCT analysis to non-binary LDPC codes are as follows.
\emph{(i)}. This formula has been shown of great potential in designing excellent irregular LDPC codes
for soft decision decoders in
\cite{Luby01IrrLdpc}, where they show that given the degree distributions, one can construct decoding graphs for any number of nodes with the correct edge fractions, under \emph{belief propagation} algorithm, by using Gallager's formula.
The designed results can be directly applied to soft decision decoders.
\emph{(ii)} For practical considerations,
this formula simplifies the analysis of convergence behavior of $q$-ary LDPC codes and
makes the design of complexity-optimized $q$-ary LDPC codes possible.
From this formula\cite{Gallager63low-densityparity-check}, it is known that for a degree-$k$ check node, the probability of either no errors or of the summation of errors is equal to 0
(mod-$q$) in one of the $k-1$ parity check sets is
\begin{equation}\label{Q_out}
  Q_{{out},k} = \frac{1+(q-1)(1-\frac{qp_{{in}}}{q-1})^{k-1}}{q},
\end{equation}
where $p_{in}$ is the input error probability of messages from a variable node to a check node.
For an irregular-check-degree
depth-one tree, define $Q_{out}$ as
\begin{eqnarray}\label{Q_out_all}
  Q_{{out}} =\sum \limits_{k \geqslant 2}\rho_k Q_{{out},k}.
\end{eqnarray}

For a variable with degree $d_v = i$, the output message error probability $p_{i,out} = f_i(p_{in})$, where $f_i$ is the uni-parametric element EXIT chart given by
\begin{multline}\label{f_out}
  f_i(p_{in}) = p_0 - p_0\sum \limits_{l=l_0} \limits^{i-1} {{i-1} \choose {l}} Q_{out}^{l}(1-Q_{out})^{i-1-l}+\\
  (1-p_0)(q-1)\sum \limits_{l=l_0} \limits^{i-1} {{i-1} \choose {l}}\left ( \frac{1-Q_{out}}{q-1} \right )^{l} \left ( 1- \frac{1-Q_{out}}{q-1} \right )^{i-1-l},
\end{multline}
where $p_0$ is the initial error probability from the channel.
The second additive term in Eq. \eqref{f_out} is
the probability of message received
in error in the variable and then corrected,
while the third additive term is the probability
that $l_0$ check nodes agree on the same error message.
$l_0$ is the smallest integer chosen to minimize $p_{out}$, subject to
$l_0>(i-1)/2$, for which
\begin{equation}
\label{choose_B}
  \frac{1-p_0}{p_0}\leqslant \frac{Q_{out}^{l_0}(q-1)^{i-2}}{(1-Q_{out})^{(2l_0 +1-i)}(q-2-Q_{out})^{(i-1-l_0)}}.
\end{equation}

From\cite{ge09NLDPC,smith10LdpcPCT}, it is known that the overall
decoding complexity is proportional to $NE$, where $N$ is the number of decoding
iterations and $E$ is the number of edges in Tanner graph. Since
each codeword encodes $Rn\log q$ information bits, the decoding complexity per information
bit is $O(\frac{NE}{Rn\log q})$. Then the decoding complexity can be formulated as
\begin{equation}
\label{decod_compl}
  K = \frac{NE}{Rn\log q} = \frac{N(1-R)}{R\log q \sum_{i \geqslant 2} \frac{\rho_i}{i}}.
\end{equation}
So, complexity optimization is equivalent to finding
the unique local minimum of $K$ in general, because the convexity can not
be always guaranteed\cite{Wei05PctB,smith10LdpcPCT}.

\subsection{A General Method for Constructing Irregular $q$-ary LDPC Codes with Optimized PCT}
\label{section_PCT}
The fact that $q$-ary LDPC codes with small mean column weight $\bar{d}_v$ can outperform
other LDPC codes, has been known for years\cite{ge09NLDPC,Davey98montecarlo}.
For large field order, average columns weight $\bar{d}_v$ of
the best $q$-ary LDPC\cite{ge09NLDPC,Davey98montecarlo} will tend to $2$,
which is also called $q$-ary cycle LDPC codes\cite{Huang10CycleNLDPC}.
Irregular $q$-ary LDPC codes with small $\bar{d}_v$, \ie $2<\bar{d}_v<3$, can outperform other LDPC codes\cite{ge09NLDPC,Davey98montecarlo}.
In this manuscript, we do not restrict the
variable degree to only two small numbers
as in \cite{jie08Underwater}, hoping to find better codes.

Considering irregular $q$-ary LDPC codes with degree distribution $\lambda(x)$ and $\rho(x)$,
we set a target rate $R_0$, $R\geqslant R_0$.
Then the optimization algorithm in\cite{smith10LdpcPCT} is modified as
\begin{eqnarray}
\label{PCT}
\nonumber
  \text{minimize} && \frac{1-R_0}{R_0\log(q) \sum \frac{\rho_i}{i}} \int_{p_t}^{p_0}{\left( p\ln \left( \frac{p}{\sum \lambda_i f_i(p)} \right)\right)^{-1}dp}.\\
\nonumber
  \text{subject to}
  && p < \sum \lambda_i f_i(p);\\ \nonumber
  && \sum_i(\lambda_i/i) \geqslant \frac{\sum_i(\rho_i/i)}{1-R_0};\\
\nonumber
  && \lambda_i\geqslant 0,\rho_i \geqslant 0;\\
\nonumber
  && \sum_{i} \lambda_i = \sum_{i} \rho_i = 1;\\
  && \|{\lambda}-\bar{ \lambda}\|_\infty<\zeta_1, \|\rho-\bar{\rho}\|_\infty<\zeta_2.
\end{eqnarray}
where $\bar{\lambda}$ and $\bar{\rho}$ can be initialized as the threshold-optimized LDPC
codes suggest \cite{ge09NLDPC,smith10LdpcPCT}.
$R_0$ is fixed which is lower than the rate of the code $(\bar{\lambda},\bar{\rho})$.
$\zeta_1$ and $\zeta_2$ are carefully set to be small values to guarantee
finding the unique local maximum \cite{Wei05PctB,smith10LdpcPCT}.
The constraint $p<\sum \lambda_i f_i(p)$ is substantial for which this optimization
algorithm is valid.

\begin{figure}[!hbtp]
\begin{center}
\includegraphics[width=3in]{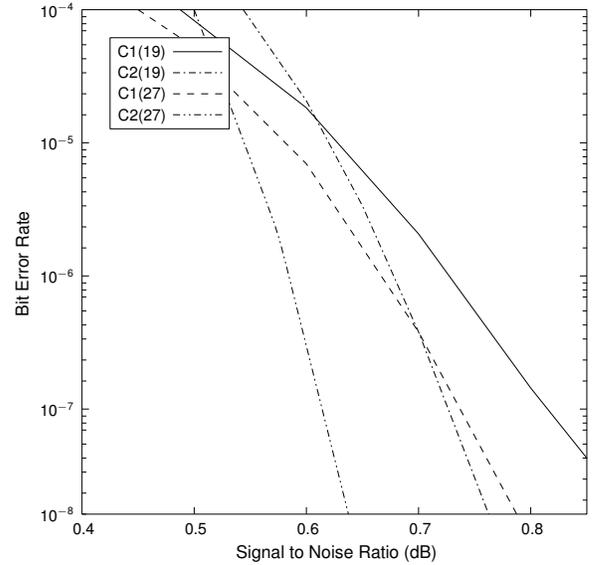}
\end{center}
\vspace*{-.3cm}
\caption{Performance comparison according to BER.}
\label{fig_pct_ber}\vspace*{-.3cm}
\end{figure}

Note that, this \emph{irregular} algorithm is different to
the \emph{quasi-regular} optimization
in\cite{smith10LdpcPCT} in the sense that the proposed algorithm updates $\bar{\lambda}$ and $\bar{\rho}$
by the recent optimal values in each iteration
through which we obtain the convergence-optimized $q$-ary LDPC codes..
More importantly, a mild condition, \ie $\{\lambda_i|f(p_{{in}})\geqslant e^2 p_{{in}} \}$, is given in\cite{Wei05PctB,smith10LdpcPCT}, under which $f(p_{{in}})$ is a convex function of $\lambda_i$. The complexity-optimized $q$-ary LDPC codes, resulting from our irregular algorithm, has a little lower threshold than the original one, but converges faster at higher SNR regime.
We take the $q$-ary LDPC codes with variable degrees restricted to $2$ and $3$
\cite{Huang10CycleNLDPC,jie08Underwater}
for example.
If the message error probability is sufficiently small, then $Q_{out,k}\approx 1-(k-1)p_{in}$. From Eq.~\eqref{Q_out_all}, calculate $Q_{out}\approx 1-(\tau_1 + \rho_{\tau_2}-1)p_{in}$, and $Q^2_{out}\approx 1-2(\tau_1+\rho_{\tau_2}-1)p_{in}$, where $\tau_1$ and $\tau_2$ is the check degrees.
In addition, the element EXIT charts of the designed $q$-ary LDPC codes are
\begin{eqnarray}
\nonumber
  f_2(p_{in}) & = & 1-(2-p_0)Q_{out},\\
\nonumber
  f_3(p_{in}) & = & p_0 + \frac{1+p_0}{q-1}(1-2Q_{out}+Q^2_{out})-Q^2_{out}.
\end{eqnarray}
Then, we have
\begin{equation}\label{p_out_approx}
  f(p_{in})\approx (p_0 - 1) + (2-\lambda_2 p_0)(\tau_1+\rho_{\tau_2}-1)p_{in}.
\end{equation}

It is easy to verify that Eq.~\eqref{p_out_approx} does not always satisfy the convex condition. Numerical simulations nevertheless suggest that, there exists a unique local optimum.
In table~\ref{tb_dv}, we give the minimum average column weight of the parity check
matrix, \ie $T_{\bar{d}_v}$, in terms of the code rate, such that the optimization algorithm is valid.

\section{Simulation Results}
\label{section_sim}
\begin{table*}[!t]
\caption{Number of iterations for Gallager decoding algorithm b.}
    \begin{center}
        \begin{tabular}[r]{|c|c|c|c|}
          \hline
          $(\bar{d}_v,\bar{d}_c)$ & $f(p_{in})$ & estimated & actual \\ \cline{1-4}
          $(2.7,3.75)$ & $0.62p_{in}+4.97p^2_{in}-18.24p^3_{in}+27.53p^4_{in}-23.28p^5_{in}+10.75p^6_{in}-2.09p^7_{in}$ & $21.1$ & $22$ \\ \hline
          $(2.7,3.6)$ & $0.59p_{in}+5.3p^2_{in}-16.25p^3_{in}+23.20p^4_{in}-18.20p^5_{in}+8.01p^6_{in}-1.45p^7_{in}$ & $19.04$ & $18$ \\ \hline
          $(2.65,3.53)$ & $0.69p_{in}+4.71p^2_{in}-14.46p^3_{in}+20.11p^4_{in}-15.53p^5_{in}+6.48p^6_{in}-1.13p^7_{in}$
          & $26.67$ & $26$ \\ \hline
          $(2.68,3.94)$ & $0.70p_{in}+5.79p^2_{in}-20.19p^3_{in}+32.23p^4_{in}-28.81p^5_{in}+14.11p^6_{in}-2.93p^7_{in}$ & $28.81$ & $28$ \\ \hline
          $(2.65,3.68)$ & $0.72p_{in}+5.00p^2_{in}-16.32p^3_{in}+24.15p^4_{in}-19.92p^5_{in}+8.95p^6_{in}-1.69p^7_{in}$ & $30.97$ & $$31$$ \\
          \hline
        \end{tabular}
    \end{center}
    \label{tb_iter}
\end{table*}

The $q$-ary LDPC codes in the manuscript are construct
by the modified progressive edge-growth (PEG) algorithm.
If the variable degrees are restricted to $2$ and $3$,
We estimate the number of iterations when the message error probability is reduced to $10^{-6}$ from $10^{-2}$. Table~\ref{tb_iter} gives the estimated and actual number of iterations according to different  $\bar{d}_v$ and $\bar{d}_c$ for Gal-b.
Table \ref{tb_dv} gives
the required smallest $\bar{d}_v$, \ie $T_{\bar{d}_v}$,
for different code rate $R$,
such that the proposed optimization algorithm is valid for the soft decision decoder.

\begin{figure}[!hbtp]
\begin{center}
\includegraphics[width=3in]{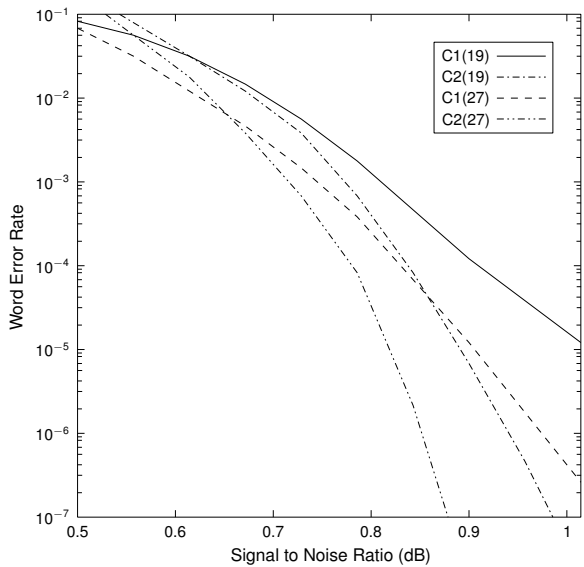}
\end{center}
\vspace*{-.3cm}
\caption{Performance comparison according to WER.}
\label{fig_pct_wer}\vspace*{-.3cm}
\end{figure}

Then, we show how to reduce the decoding complexity
of a given code.
Considering the threshold optimized $4$-ary LDPC codes with block length 30000 bits
reported in \cite{ge09NLDPC,smith10LdpcPCT},
characterized by
$\lambda(x)=0.249009x+0.200042x^2+0.02177703x^5+0.161403x^6+0.0489424x^8+
0.0381342x^{16}+0.0874772x^{18}+0.0154621x^{19}+0.177761x^{49}$ and
$\rho(x)=0.439929x^7+0.560007x^8$,
the complexity optimized $4$-ary LDPC code characterized by
$\lambda(x)=0.5503x+0.0297x^3+0.1304x^4+0.2003x^{15}+0.0893x^{20}$ and
$\rho(x)=0.2998x^3+0.7002x^4$.
We give the bit error rate (BER) and word error rate (WER) in Fig.~\ref{fig_pct_ber}
and Fig.~\ref{fig_pct_wer} by calculating the average error rate from 100 times experiments.
We expect that the complexity optimized code will
reach a BER of $10^{-7}$ faster at a smaller number of iterations,
while maintaining the excellent performance as the original one.
Let $C_1(N)$ and
$C_2(N)$ be the original and optimized codes, respectively, where
$N$ is the number of iterations.
Fig.~\ref{fig_pct_ber} shows that
the optimized code
outperforms the original one with faster convergence rate
at a small $N$.
$C_2(19)$ even converges faster than $C_1(27)$.
The convergence process has been accelerated by $30\%$
regarding the number of decoding iterations.
\begin{table}[!htp]
    \caption{The smallest $\bar{d}_v$ required for different rates}
    \begin{center}
        \begin{tabular}[r]{|c|c|c|c|c|c|c|}
          \hline
          $R$ & 1/6 & 1/5 & 1/4 & 1/3 & 1/2 & 2/3 \\ \cline{1-7}
          $T_{\bar{d}_v}$ & 2.37 & 2.40 & 2.48 & 2.56 & 2.70 & 2.81 \\
          \hline
        \end{tabular}
    \end{center}
    \label{tb_dv}
\end{table}

\section{Conclusion and Discussions}
The proposed PCT algorithm is used to design irregular nonbinary LDPC codes with optimized decoding complexity. However, the encoding complexity is not optimized during the design procedure.
A future work of this manuscript is to construct structured nonbinary LDPC codes
that can achieve optimized decoding complexity and optimized encoding complexity at the same time.
In addition, upper bounds of message error probability are used to analyze the performance of nonbinary LDPC codes, which results in an approximated PCT analysis for the soft decision decoder.
In order to achieve faster convergence performance, we need to construct more accurate PCT algorithms.
Another future work of this manuscript is to find more accurate uni-parametric representation of the decoding trajectory for the nonbinary soft decision decoders.

\ifCLASSOPTIONcaptionsoff
  \newpage
\fi

\bibliographystyle{IEEEtran}
\bibliography{IEEEabrv,QLDPC}


%

\end{document}